\newcommand{\PaperTitle}{ReACKed QUICer: Measuring the Performance of Instant Acknowledgments in QUIC Handshakes}

\documentclass[sigconf]{acmart}

\usepackage{graphicx}
\usepackage[labelformat=simple]{subcaption}

\usepackage[]{todonotes} %
\usepackage{multirow}
\usepackage{balance}

\usepackage{dblfloatfix}
\usepackage[absolute,showboxes]{textpos}


\newcommand{\result}[1]{}

\definecolor{myred}{cmyk}{0, 0.7808, 0.4429, 0.1412}

\newcommand{\done}[1]{}

\usepackage{pifont}

\usepackage{xspace}
\newcommand{\etal}{\textit{et al.}~}
\newcommand{\eg}{\textit{e.g.,}\xspace}
\newcommand{\ie}{\textit{i.e.,}~}
\newcommand{\cf}{\textit{cf.,}~}

\newcommand{\one}{({\em i})\xspace}
\newcommand{\two}{({\em ii})\xspace}
\newcommand{\three}{({\em iii})\xspace}

\makeatletter
\renewcommand{\paragraph}[1]{\vspace*{0.03in}\noindent{\bf #1.}\hspace{0.25ex \@plus1ex \@minus.2ex}}
\newcommand{\paragraphNoDot}[1]{\vspace*{0.03in}\noindent{\bf #1}\hspace{0.25ex \@plus1ex \@minus.2ex}}
\makeatother

\copyrightyear{2024}
\acmYear{2024}
\setcopyright{rightsretained}
\acmConference[IMC '24]{Proceedings of the 2024 ACM Internet Measurement
Conference}{November 4--6, 2024}{Madrid, Spain}
\acmBooktitle{Proceedings of the 2024 ACM Internet Measurement Conference (IMC
'24), November 4--6, 2024, Madrid, Spain}\acmDOI{10.1145/3646547.3689022}
\acmISBN{979-8-4007-0592-2/24/11}

\makeatletter
\gdef\@copyrightpermission{
  \begin{minipage}{0.3\columnwidth}
   \href{https://creativecommons.org/licenses/by/4.0/}{\includegraphics[width=
0.90\textwidth]{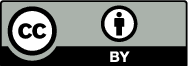}}
  \end{minipage}\hfill
  \begin{minipage}{0.7\columnwidth}
   \href{https://creativecommons.org/licenses/by/4.0/}{This work is licensed
under a Creative Commons Attribution International 4.0 License.}
  \end{minipage}
  \vspace{5pt}
}
\makeatother

\begin{document}

\title{\PaperTitle}

\author{Jonas M\"ucke}
\email{jonas.muecke@tu-dresden.de}
\affiliation{%
  \institution{TU Dresden}
  \city{Dresden}
  \country{Germany}
}

\author{Marcin Nawrocki}
\email{marcin.nawrocki@netscout.com}
\affiliation{%
  \institution{NETSCOUT}
  \city{Westford}
  \state{MA}
  \country{USA}
}

\author{Raphael Hiesgen}
\email{raphael.hiesgen@haw-hamburg.de}
\affiliation{%
  \institution{HAW Hamburg}
  \city{Hamburg}
  \country{Germany}  
}

\author{Thomas C. Schmidt}
\email{t.schmidt@haw-hamburg.de}
\affiliation{%
  \institution{HAW Hamburg}
  \city{Hamburg}
  \country{Germany}
}

\author{Matthias W\"ahlisch}
\email{m.waehlisch@tu-dresden.de}
\affiliation{%
  \institution{TU Dresden}
  \city{Dresden}
  \country{Germany}
}

\begin{abstract}

In this paper, we present a detailed performance analysis of QUIC instant~ACK, a standard-compliant approach to reduce waiting times during the QUIC connection setup in common CDN deployments.
To understand the root causes of the performance properties, we combine numerical analysis and the emulation of eight QUIC implementations using the QUIC Interop Runner.
Our experiments comprehensively cover packet loss and non-loss scenarios, different round trip times, and TLS certificate sizes.
To clarify instant ACK deployments in the wild, we conduct active measurements of 1M~popular domain names.
For almost all domain names under control of Cloudflare, Cloudflare uses instant ACK, which in fact improves performance.
We also find, however, that instant ACK may lead to unnecessary retransmissions or longer waiting times under some network conditions, raising awareness of drawbacks of instant ACK in the future.
\end{abstract}

\maketitle

\definecolor{boxgray}{rgb}{0.93,0.93,0.93}
 \textblockcolor{boxgray}
 \setlength{\TPboxrulesize}{0.7pt}
 \setlength{\TPHorizModule}{\paperwidth}
 \setlength{\TPVertModule}{\paperheight}
 \TPMargin{5pt}
 \begin{textblock}{0.8}(0.1,0.04)
   \noindent
   \footnotesize
   If you refer to this paper, please cite the peer-reviewed publication: Jonas M\"ucke, Marcin Nawrocki, Raphael Hiesgen, Thomas C. Schmidt, Matthias W\"ahlisch. 2024. ReACKed QUICer: Measuring the Performance of Instant Acknowledgments in QUIC Handshakes. 
   In \emph{Proceedings of ACM Internet Measurement Conference (IMC)}. ACM, New York. \url{https://doi.org/10.1145/3646547.3689022}
\end{textblock}

\section{Introduction}
\label{sec:introduction}

Low latencies are key for Content Delivery Networks~(CDNs), even if savings are in the range of few milliseconds.
With handshakes often spanning multiple round trip times (RTTs) between client and CDN server, a reduction of this latency is beneficial.
QUIC~\cite{RFC-9000} promises such a reduction by combining the transport and encryption (TLS) handshakes to establish a connection in a single round trip time (1-RTT).
This makes QUIC popular in large-scale content provider setups~\cite{rpdh-flaqi-18,ss-athp-15,zbsja-io9ae-21} and recent protocols, \eg HTTP/3 \cite{RFC-9114}.

Common CDN deployments challenge a fast connection setup in QUIC.
During the connection setup in QUIC,  the server requires a TLS certificate to answer the client connection request. 
In many CDN setups, the server needs to fetch this certificate from a backend server.
Waiting until the certificate is available inflates the RTT.
Alternatively, the server could acknowledge the client request instantly and continue the handshake after the certificate is received. 
We dub this behavior \emph{instant ACK}.
An instant ACK promises a more realistic RTT estimation between server and client, and therefore leads to more realistic timeouts in case of packet losses.
Cloudflare, a major CDN, implements this behavior \cite{nthms-ibtcq-22}.
Whether an instant ACK is an advantageous option during the QUIC connection setup is an open topic, though.

In this paper, we empirically analyze the performance of instant ACK in multiple scenarios and compare with deployments in which a server waits until the certificate is available.
Our analysis allows us to conclude when an instant ACK is advantageous, how much improvement can be achieved in real deployments, and whether further adoption of instant ACK would be beneficial to reduce connection setup latencies.
Furthermore, we study the adoption of instant ACK in CDNs and measure improvements within one major CDN.
In detail, our contributions are as follows:
\begin{enumerate}
    \item We analyze the effects of instant ACK, finding improvements and deterioration. 
    Probe Timeouts (PTOs) are improved by $3\times$ the delay between frontend server and certificate store.
    On packet loss, recovery is faster -- in the order of 10 ms in the wild.
    When QUIC servers are blocked due to the anti-amplification limit, instant ACK may improve handshake completion by up to 1~RTT.
    \item We find that  the instant ACK induces spurious retransmissions in scenarios of high latency between frontend server and certificate store.
    \item We measure the global deployment of instant ACK. Up to 99.9 \% of Cloudflare, 41~\% of Amazon, 32.2 \% of Akamai, and 11.5~\% of Google domains on the Tranco Top 1M use instant~ACK.
    \item We analyze in which scenarios QUIC connections benefit most from the instant ACK optimization and show that the Cloudflare deployment lies within that~range.
    We identify tuning parameters for client and server configurations.
    \item We conclude from comparing different implementations and from numerical analysis  that improvements are not implementation dependent but on the protocol level. We also find inconsistent client behavior and work with the maintainers on  resolving this.
\end{enumerate}

\section{Background and Related Work}
\label{sec:background}
\begin{figure}
    \begin{subfigure}[t]{0.49\linewidth}
    \centering
    \includegraphics[width=1.0\columnwidth]{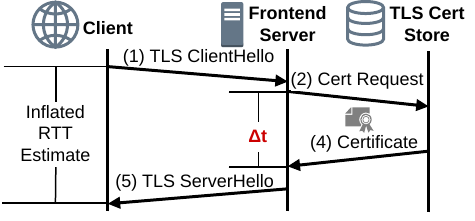}
    \caption{Wait for certificate (WFC)}
    \label{fig:WFC-timeline}
    \end{subfigure}
    \begin{subfigure}[t]{0.49\linewidth}
    \centering
    \includegraphics[width=1.0\columnwidth]{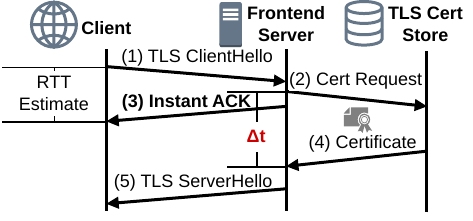}
    \caption{Instant ACK (IACK)} %
    \label{fig:IACK-timeline}
    \end{subfigure}
    \caption{QUIC server behaviors in common CDN deployments. The delay $\Delta t$ between frontend server and certificate store inflates the RTT estimate for subsequent data from the frontend server, if the frontend server does not use IACK.}
    \vspace{-0.5cm}
    \label{fig:IACK-WFC-timeline}
\end{figure}

\autoref{fig:IACK-WFC-timeline} illustrates a simplified QUIC handshake in common CDN~deployments.
In response to the TLS ClientHello, the server needs encryption keys to compile the TLS ServerHello, certificate and signatures, and returns them to the client.
Large CDNs, such as Akamai and Cloudflare, separate TLS endpoints and certificate stores~\cite{cassb-sdarm-20,ss-athp-15} to liberate TLS endpoints from storing all customer certificates.
Certificates are provisioned on the frontend server when needed, often after a connection request.
The frontend server then has two options.
\one The server waits until the certificate is retrieved from the certificate store, we dub this \emph{wait for certificate~(WFC)}, or \two replies immediately with an ACK to the client and delivers the ServerHello as soon as the certificate is available, we dub this \emph{instant ACK (IACK)}.

\paragraph{Calculation of the Probe Timeout}
The RTT estimation determines the Probe Timeout (PTO).
The PTO is initialized with $3\times$ of the first RTT~sample.
Although ACKs include an acknowledgment delay to correct RTT estimation of intentionally delayed ACKs, the PTO initialization disregards this delay.
Therefore, the only option to provide the client with an accurate PTO is via the instant ACK.
Subsequently, the PTO is based on the exponentially weighted moving averages of the smoothed RTT plus RTT variation, and the maximum acknowledgment delay provided by each endpoint.
A PTO timer is set and reset when a packet eliciting an ACK is sent or acknowledged and when Initial or Handshake keys are discarded.
Whenever a PTO expires the sender must send 1-2~datagram probes, which elicit an ACK and exponentially backoff the PTO.
Retransmitting tail bytes in-flight is recommended;  they can be bundled with outstanding data if available. 
If neither are available, senders will fall back to sending PING frames, which do not carry further information.
Sending acknowledgments, \eg in response to PING frames, does not cause a PTO reset.

\paragraph{Potential disadvantages of wait for certificate}
When a server waits for the certificate before responding to a client, the handshake is delayed, see $\Delta t$ in \autoref{fig:WFC-timeline}.
This delay impacts the Round Trip Time~(RTT) estimation by the client.
The client then assumes an RTT that is inflated compared to the RTT between the client and frontend server, which delivers the web content.

\begin{figure}
    \centering
    \includegraphics[width=\columnwidth]{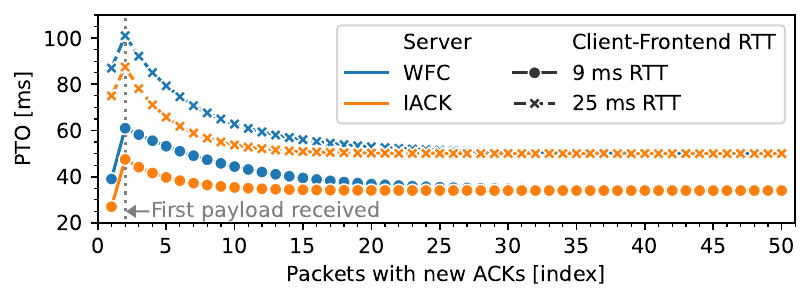}
    \caption{Calculated evolution of the Probe Timeout (PTO) assuming that all subsequent packets arrive exactly after one RTT and the instant ACK is delivered 4~ms earlier. 
    The instant ACK leads to a PTO improvement of $3\times \Delta t$.}
    \label{fig:PTO-development}
    \vspace{-1cm}
\end{figure}

\paragraph{Potential advantages of instant ACK}
The instant ACK reduces the PTO and consequently the time until retransmissions occur.
This is important \one under packet loss and \two during the handshake, when the server is limited by the QUIC anti-amplification limit.
We illustrate the evolution of the PTO in a static setting in \autoref{fig:PTO-development}. 
An inflated PTO slowly approaches the actual PTO between client and frontend server, resulting in negative effects for content delivery in case of packet loss.
However, when the delay between the frontend server and certificate store is larger than the first instant ACK PTO, this leads to spurious~retransmits.

To avoid amplification attacks, the server is limited to send $3\times$ the data received from the client until the client address is verified~\cite{RFC-9000}.
If the handshake exceeds this limit, the server needs to wait for additional client data to increase its amplification budget.
In this situation, earlier PTO expiration leads to earlier probe packets, which accelerate the QUIC handshake even without packet loss.
Earlier work showed that Cloudflare implements instant ACKs~\cite{nthms-ibtcq-22}, but we also find that other CDNs experiment with instant ACKs.

\paragraph{Related work}
Analyzing QUIC~performance is an ongoing research field.
Flow control, multiplexing strategies~\cite{mhlq-ssdds-20,yb-dppq-21}, robustness~\cite{nhsw-qqqrs-21}, resource prioritization~\cite{skw-airph-22}, CPU~\cite{jzkpc-qheph-23} or NIC~\cite{yeous-mqqwn-20} architectures, and the congestion control (CC) algorithm~\cite{mmm-oeaqc-21,mll-usqcc-22,hgfhb-pqcca-22,wrwh-ppowo-19} impact performance.
Short-lived connections benefit most from QUIC~\cite{yb-dppq-21,sppb-eqpow-22}, and QUIC shows better performance~\cite{sppb-eqpow-21,slm-ebqeb-21,wrwh-ppowo-19,yb-dppq-21} in the presence of lossy links. %
Compared to TCP, QUIC improves RTT estimation~\cite{sppb-eqpow-22,yb-dppq-21} due to simplified packet numbering.
Default host buffer sizes, however, are too small by an order of magnitude~\cite{jzkpc-qheph-23} to achieve high goodput, and current TLS~certificate configurations prevent 1-RTT handshakes~\cite{nthms-ibtcq-22}.
Prior work relies on random packet drop rates or selected deployments, including IoT scenarios~\cite{dj-ebhpi-21}, satellite communication~\cite{kcbo-epqhs-22,hdfhh-ratpp-22}, and cellular networks~\cite{mmm-oeaqc-21,hgfhb-pqcca-22}, as well as interoperability testing~\cite{si-aqit-20}. 
The combination of client and server implementations lead to different performance expectations~\cite{jzkpc-qheph-23}.

We extend the field by analyzing the performance impact of instant ACK in a well-known testbed and a common CDN setup.
Our emulation comprises multiple QUIC implementations, RTTs ranging from local to long distance, and distinct packet drops instead of relying on stochastic or exemplary deployments. 
This allows pinpointing the effects of instant ACK.

\section{Measurement Method and Setup}
\label{sec:method}

To compare the impact of instant ACK on QUIC handshakes under conditions of packet loss, detailed information on the exchanged packets, their timing, and the experienced packet loss are required. 
We rely on a controlled testbed to obtain this information.
We measure instant ACK deployments by performing QUIC handshakes with several servers in the wild and analyzing response traffic. 
Based on the collected data we calculate PTOs and compare behaviors.

\paragraph{Microscopic view---Testbed}
We emulate network conditions using the QUIC Interop Runner (QIR), a container-based framework for interoperability testing of QUIC implementations \cite{si-aqit-20}.
QIR captures packets and collects Qlog information from clients and servers. %
Qlog~\cite{draft-ietf-quic-qlog-main-schema}, a structured logging format for QUIC, contains data about sent packets, received packets, and \textit{recovery:metrics}, including the smoothed RTT and RTT variation calculated by the implementation.
Nonetheless, implementations differ in how often and how exhaustive \textit{recovery:metrics} are exposed (see~\autoref{app:client-implementation}).
Writing Qlogs might slow down implementations \cite{jzkpc-qheph-23}.
The exposed information, however, is the best approximation of a systems view on connections since delays cover the networking stack, scheduling, and abstraction layers---similar to real-world deployments and more detailed than packet captures.
Qlog is commonly used for such analyses~\cite{mlrpq-tqd-18,mhlq-ssdds-20,yb-dppq-21}.
Multiple repetitions help us to identify patterns.
To ensure consistency, we calculate PTOs based on sent and received packets according to the standard~\cite{RFC-9002}.

Our evaluations consist of an HTTP \texttt{GET} request and response.
We test the transfer of 10 KB and 10 MB randomly generated files, as representatives for an embedded library and larger websites.

The server is armed with two certificates: one allows a 1-RTT handshake~(1,212~B), the other~(5,113~B) exceeds the anti-amplification limit.
Backend--frontend delays are emulated by a configurable sleep period in the server code.
All simulations use a bandwidth of 10~Mbit/s.
RTTs are composed of symmetric one way delays between 0.5~ms and 150~ms, based on typical RTTs measured in different regions of the world \cite{mhrr-rttdi-23}.
Every test is repeated 100~times.

\begin{figure}
    \centering
    \includegraphics[width=\columnwidth]{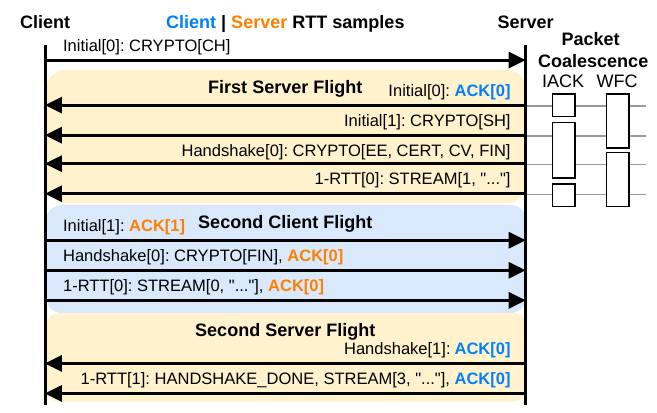}
    \caption{QUIC 1-RTT connection setup~\cite{RFC-9000} and coalescence differences between IACK and WFC. 
    Blue and orange ACKs indicate client and server RTT samples.}
    \label{fig:quic-connection-setup}
    \vspace{-1.0cm}
\end{figure}

Packet loss is often modeled as a random share of lost packets or observed on real world links.
Our emulation instead simulates particular datagram losses to better understand root causes of specific performance properties.
The QUIC protocol allows coalescence of multiple QUIC packets into a single UDP datagram, \ie an entire flight can be transmitted in one datagram (see \autoref{fig:quic-connection-setup}).
Implementations use this feature to different extents. %
Consequently, the impact of a single packet loss differs between implementations.
Unless stated otherwise, we match lost datagrams to their QUIC content and compare equal information loss. 
This ensures that different packet coalescence does not influence protocol level conclusions.

We perform QUIC connections from eight client implementations (\textit{aioquic, go-x-net, mvfst, neqo, ngtcp2, picoquic, quic-go, quiche}) to a \textit{quic-go} server modified to support IACK.
Clients request data via HTTP/1.1 and HTTP/3 (except \textit{go-x-net}, which does not implement HTTP/3).
Although Cloudflare is reported to implement instant ACK, the public source code for \textit{quiche} does not contain IACK.
The simulations are distributed over 18 test runners (virtual machines with 4 AMD EPYC 7513 Cores, 4 GB RAM, Fedora 39).
Virtualization and containerization are a common practice in industry. 
To ensure that these abstractions do not bias our results, we verify them by thoroughly inspecting packet captures and tracing differences back to the protocol behavior.

\paragraph{Macroscopic view---Instant ACK deployments}
We perform QUIC handshakes and HTTP/3 \texttt{HEAD} requests using QScanner~\cite{zbsja-io9ae-21}. 
We target 1M~domain names in the Tranco toplist~\cite{pgtkw-trots-19} from August~06,~2024 and repeat our measurements on three subsequent days.
We then map the contacted IP addresses to ASes and on-net CDN deployments (see \autoref{app:macroscopic-view}).

To consider side effects introduced by third parties (\eg caching of certificates on frontend servers), we add twelve of our own otherwise unused domains to the Cloudflare Free Tier CDN with default settings. 
The TLS certificates of those domains are managed by Cloudflare and signed by Google Trust Services.
For comparison, we select six domains from the Top 1000 Tranco domains that are served by Cloudflare.
We then schedule one connection per minute to six of our domains and the six domains selected from Tranco. 
To our remaining six domains, we schedule 60 connections per minute. 
We run these measurements for one week, collect all response traffic, and analyze the content using a packet dissector.
The location of the replying server is derived from the IATA location identifier included in the HTTP header \texttt{Cf-Ray}, which is provided by Cloudflare servers.
We only consider responses from servers located in the same city as our measurement probe to eliminate larger network effects and ensure that we reach the same cluster.
Additionally, we account for packet loss by including only responses that contained the first ACK of the connection.
We compare the request and arrival times of the received ACK, ServerHello, and coalesced ACK--ServerHello.

To cross-verify our results, we perform all measurements from a European university network~(Hamburg, DE) and Google Cloud VMs in North America~(Los Angeles, US), South America~(Sao Paulo, BR), and Asia~(Hong Kong, HK) in August~2024.

\section{Results}
\label{sec:analysis_qir}

In this section, we \one assess when an instant ACK yields most benefits, \two measure the performance impact of instant ACK in QUIC implementations with and without packet loss, and \three analyze instant ACK in global CDN~deployments.

\subsection{Baseline Performance}
\label{subsec:analysis-baseline}

\begin{figure}
    \includegraphics[width=\columnwidth]{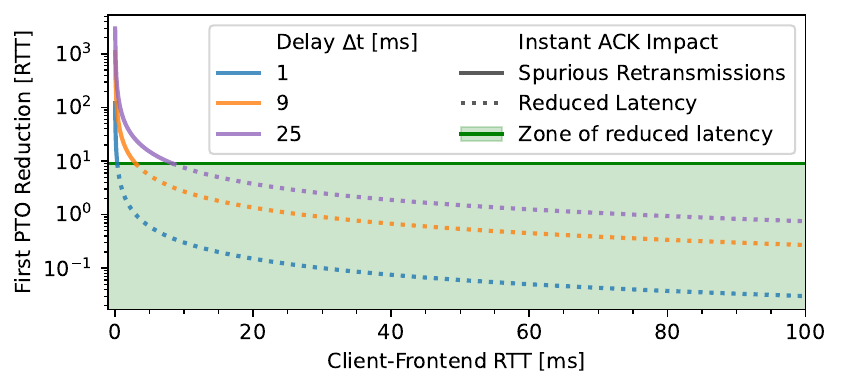}
    \caption{First PTO improvement according to RFC9002~\cite{RFC-9002}. 
    Spurious retransmits happen if the delay between Frontend Server and Cert Store ($\Delta t$) is larger than the PTO set by the client.
    Relative to the RTT, lower latency connections profit more from PTO improvement with IACK.}
    \label{fig:theoretic-improvement-factors}
  \vspace{-0.5cm}
\end{figure}

\paragraph{Numerical sweet spot analysis}
Before we present results based on our testbed, we use a numerical analysis to highlight the principle impact of instant ACK.
\autoref{fig:theoretic-improvement-factors} shows the theoretical PTO reduction relative to the RTT and the area in which no spurious retransmissions occur for different $\Delta t$ and varying RTTs.

Instant ACK improves performance in two scenarios.
First, in case of client packet loss, the shorter PTO is beneficial because the client repeats its requests earlier. 
The maximum PTO improvement is achieved when $\Delta t$ approaches the client PTO.
Second, when $\Delta t$ exceeds the client PTO and the server is blocked by the anti-amplification limit, the client will send probe packets earlier, increasing the amplification budget leading to a faster termination of the handshake.
If the server is not blocked during the handshake instant ACK might harm because the client probing due to expired PTO causes additional load on the server without any~benefit.

\paragraph{QUIC stack delays}
The $\Delta t$ does not only consist of the network delay between the frontend server and certificate store but also of the local processing time.
In our testbed, we measured that the delay difference between IACK and WFC reported by the different QUIC clients is on median 2.9~ms~(\textit{neqo}) to 7.8~ms~(\textit{mvfst}), except for \textit{go-x-net}.
\textit{go-x-net} introduces high variations in individual measurements (median 0.1~ms to 12.7~ms) and partly reports erroneous values.

At the server side,  to send an instant ACK, the QUIC stack needs to calculate the Initial keys based on information from the client packet.
In case of a coalesced message, additionally, handshake keys need to be calculated and TLS ServerHello, certificate, and signatures compiled.
Profiling 100,000 handshakes of our server implementation \textit{quic-go} reveals that signature calculation is the single most CPU consuming function.

Unless specified otherwise, we do not add delays in our emulations.

\paragraph{Instant ACK improves the first PTO}
Our measurements confirm that instant IACK improvements are independent of the RTT for all client implementations.
Only if the default PTO is lower than the RTT, side effects can occur (see \autoref{app:client-implementation}).
In a network with lower latencies, the same improvement would be relatively more beneficial (see \autoref{fig:theoretic-improvement-factors}).
This aligns with the theoretically possible improvement, although accuracy in practice is limited by the information exposed in Qlog (see~\autoref{app:client-implementation}). 
Accounting for the influence of the default PTO and erroneous PTO calculation, the median PTO improvement between IACK and WFC for 100~measurements per RTT (1~ms \dots 300~ms) ranges between 7~ms and 24.7~ms.
Given that RTTs between end-users and CDN servers are typically in the same range~\cite{tgdmm-fyaew-18} such an improvement is significant.

\begin{figure}[]
\begin{subfigure}[t]{0.49\textwidth}
    \centering
    \includegraphics[width=\linewidth]{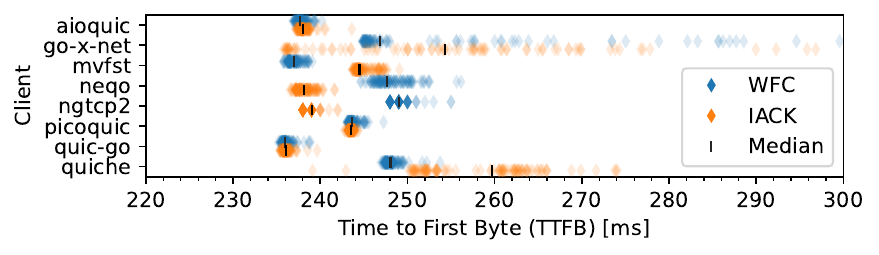}
    \caption{HTTP/1.1}
    \label{fig:ttfb-amplification-limit-goodput}
\end{subfigure}
\begin{subfigure}[t]{0.49\textwidth}
    \centering
    \includegraphics[width=\linewidth]{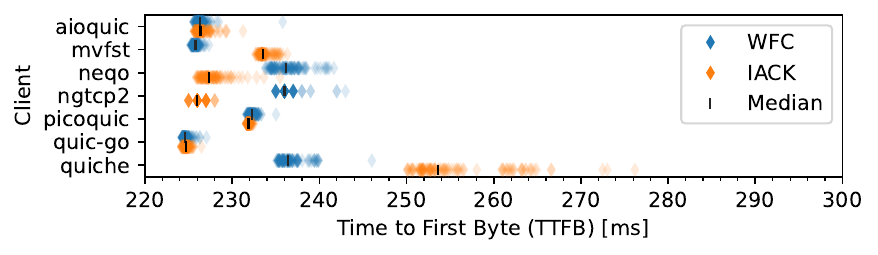}
    \caption{HTTP/3}
    \label{fig:ttfb-amplification-limit-http3}
\end{subfigure}
\caption{Time to First Byte (TTFB) of 10~KB file transfer at 9~ms RTT with large certificate, $\Delta t = 200~ms$, and without packet loss. Each diamond symbol represents the TTFB for an individual measurement. Intense colors indicate a concentration of values. HTTP/3 generally has a lower TTFB because the first STREAM frame received from the server is the Control Stream with the SETTINGS frame, which is send by the server immediately after the handshake completes. Compared to HTTP/1.1, this is one RTT faster.}
  \label{fig:ttfb-amplification-limit}
\end{figure}

\paragraph{Time to First Byte (TTFB) is reduced if the server is blocked by the anti-amplification limit}
When the TLS certificate size exceeds the anti-amplification limit the server must wait for additional packets from the client (\eg because the PTO expires).
The client PTO includes the client-frontend RTT. %
In a lossless scenario, the next packet will then be delivered later than $3\times RTT$.
We emulate this case using a certificate (size 5113~B) that exceeds the anti-amplification limit and a 200~ms delay before sending the ServerHello, \ie the first server packet in WFC and the second one in IACK.
\autoref{fig:ttfb-amplification-limit} shows a decreased median TTFB with most significant improvements for \textit{neqo} (9.6~ms) and \textit{ngtcp2} (10~ms). 
The server logs confirm that WFC is more likely to cause the server to block due to the anti-amplification limit, which happens significantly less when instant ACK is enabled.
In \textit{go-x-net}, the client partially initializes the smoothed RTT and RTT variation incorrectly, which leads to worse performance (\eg reported RTT 33~ms, but smoothed RTT is initialized at 90~ms). 
However, if the RTT estimates are initialized correctly, IACK improves the TTFB.
In \textit{mvfst and picoquic}, receiving an instant ACK does not cause the client to send probe packets.
For picoquic, the default server PTO is reached first, leading to equal performance of IACK and WFC.
In \textit{aioquic}, \textit{mvfst}, and \textit{quic-go}, the default client PTO expires before reception of the coalesced ACK--ServerHello and increases the amplification budget in both WFC and IACK before the server sends the ServerHello.
Using \textit{quiche}, we observe negative effects when IACK is enabled because the \emph{quiche} client drops replies to PING frames as invalid together with coalesced packets.
This then requires retransmission of the dropped information.

\subsection{Packet Loss Scenarios}
\label{subsec:analysis-loss5}

\begin{figure}
   \includegraphics[width=\linewidth]{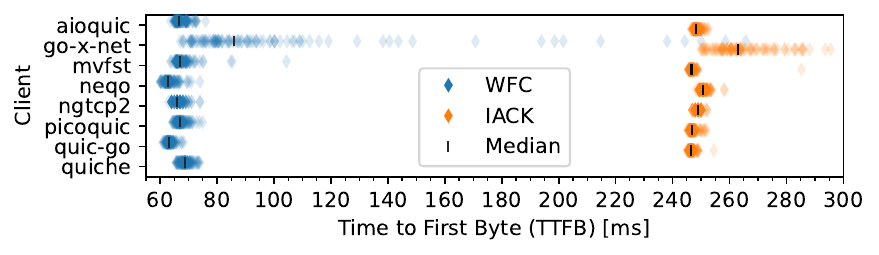}
    \caption{Time to First Byte of 10~KB file transfer at 9ms RTT under loss of packets 2 and 3~(IACK) and packet 2~(WFC) sent by the server. 
    IACK prolongs the TTFB.}
    \label{fig:ttfb-increased-iack}
\end{figure}

\paragraph{Time to First Byte is increased when the remaining first server flight is lost}
We now simulate loss of the second and third UDP datagram (IACK) and loss of the second UDP datagram (WFC) in the first server flight (\cf \autoref{fig:quic-connection-setup}).
In this scenario, WFC outperforms IACK (see \autoref{fig:ttfb-increased-iack}).
IACK requires between $\approx$177~ms (\textit{go-x-net}) and 188~ms (\textit{neqo}) more time until the first payload data is received.
This might appear surprising but relates to the QUIC protocol mechanics.
Not every ACK leads to a new RTT sample at the receiving side---only ACKs to ACK-eliciting packets do.
The IACK sent by the server is not ACK-eliciting.
In consequence, when the server receives the probe packet from the client, the server will not update its RTT estimate but relies on the default PTO.
In WFC, the server is able to update an RTT estimate, because the first server packet combines ACK--ServerHello and a ServerHello must be acknowledged by the client.
Since the PTO based on the RTT estimate is smaller than the default PTO of the server, the retransmit and thus finishing the handshake is faster in WFC than in IACK.

Our results show that \emph{quiche} does not respond to an IACK.
This is due to an uncommon behavior of \emph{quiche}: it drops connections when the same connection~ID is retired multiple times.
In our HTTP/3 measurements, we do not encounter this case.
Here, \textit{quiche} behaves like all other implementations (see \autoref{app:interop-multiple-rtts}).

\begin{figure}
    \centering
        \includegraphics[width=\linewidth]{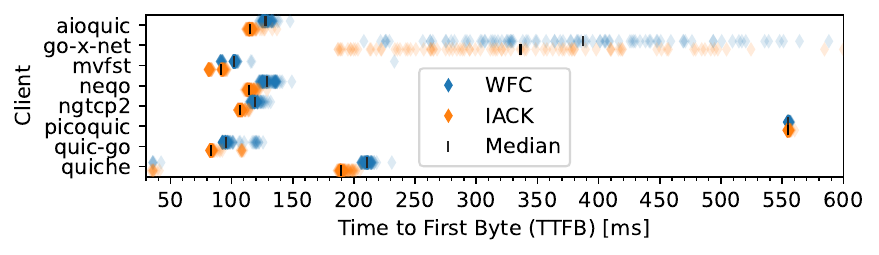}
    \caption{Time to First Byte of 10~KB file transfer at 9 ms RTT under loss of the entire second client flight (see \autoref{app:client-implementation}).   
Instant ACK improves the TTFB.
}
\label{fig:ttfb-decreased-iack}
\end{figure}

\paragraph{Time to First Byte is reduced when the second client flight is lost}
\autoref{fig:ttfb-decreased-iack} compares TTFB between WFC and IACK when the second client flight is lost (\cf \autoref{fig:quic-connection-setup}). 
The second client flight includes the first payload data a client can send in a 1-RTT handshake. 
This payload includes the HTTP request. 
If this packet is lost, a smaller PTO due to instant ACK will allow the client to resend the lost information sooner, leading to an improved TTFB. 
Independent of the simulated RTT in lossy scenarios, the first byte is received significantly sooner with on median 10~ms (\textit{mvfst}), 11~ms (\textit{aioquic, quic-go}), 12~ms (\textit{neqo, ngtcp2}), 23~ms (\textit{quiche}), and 28~ms (\textit{go-x-net}).
\textit{picoquic} does not benefit from an earlier retransmission because it ignores the lower RTT induced by IACK.

This confirms that the improvement of the instant ACK is constant across RTTs, but the relative impact is larger for short RTTs, \eg \textit{quiche} improved by 2.3 RTTs under 9 ms network latency.

\subsection{CDN Deployment}
We attempt to establish QUIC connections to all domains in the Tranco Top~1M list.
If successful, we check for instant ACK behavior, \ie whether the ClientHello is followed by a separate (server) ACK preceding the TLS ServerHello.

\begin{figure}
\includegraphics[width=\columnwidth]{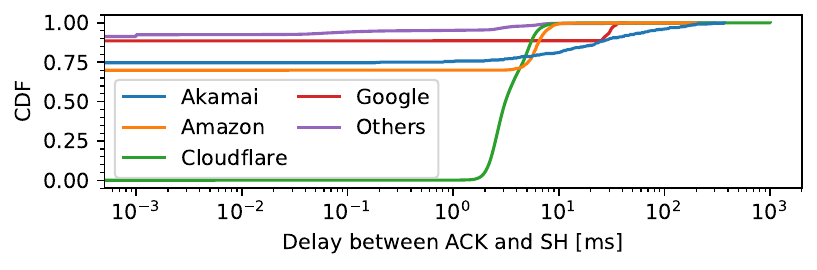}
\caption{Delay between reception of the first ACK and subsequent ServerHello (SH) from our vantage point in Sao Paulo. Coalesced ACK--SH is shown as 0 delay. Akamai is significantly slower than other CDNs to deliver the ServerHello. }
\label{fig:CDF-ACK-SH-Delay}
\end{figure}

\paragraph{Several CDNs support IACK}
Fastly, Meta, and Microsoft do not deploy instant ACK but the majority of domains (99.9\%) operated by Cloudflare do (see \autoref{tab:cdn-toplist-observations}).
Akamai, Amazon, and Google also deploy IACK but at relatively smaller scale (32.2\%, 41.0\%, and 11.5\% of the domains).
We observe the highest variations in IACK deployment from Amazon, with 8.3\% difference across measurements at the same location and 18.0\% across all measurements and vantage~points. 
\autoref{fig:CDF-ACK-SH-Delay} shows the delay between ACK and ServerHello.
On median, the IACKs arrive 3.2~ms (Cloudflare), 6.4~ms (Amazon), 30.3~ms (Google), and 20.9~ms (Akamai) earlier than the ServerHellos across all vantage~points.

\begin{table}
\caption{Domains from the Tranco Top 1M hosted by CDNs, share of instant ACK deployment, and maximum difference between measurements.
Deployment share and maximum variation are aggregated across vantage points and repetitions. Group ``Others'' represents hosting services.}
\label{tab:cdn-toplist-observations}
\begin{tabular}{lrrr}
\toprule
& & \multicolumn{2}{c}{IACK deployment enabled (max.)}
\\
\cmidrule{3-4}
CDN & Domains [\#] & Domains [\%] & Variation [\%]
\\
\midrule
Akamai & 533 & 32.2 &  12.9 
\\
Amazon & 4338 & 41.0 & 18.0 
\\
Cloudflare & 247407 & 99.9 & 0.1 
\\
Fastly & 3960 & 0.0 & 0.0 
\\
Google & 6062 & 11.5 & 11.5
\\
Meta & 112 & 0.0 & 0.0 
\\
Microsoft & 34 & 0.0 & 0.0 
\\
Others & 26404 & 21.5 & 2.3 
\\
\bottomrule
\end{tabular}
\end{table}

\paragraph{The instant ACK deployment at Cloudflare}
We now focus on Cloudflare since they have the largest deployment of instant ACK enabled services. 
To understand the impact of potential certificate caching on frontend servers, we compare handshake behavior of domains all served by Cloudflare but with different popularity (setup see \autoref{sec:method}).
Among the selected Tranco domains, we observe coalesced ACKs and ServerHellos from \textit{discord.com (91.9\% of the responses), cloudflare.com (50.5\%), tinyurl.com (17.7\%)}, and \textit{docker.com (0.7\%)}.
From \textit{udemy.com} and \textit{kickstarter.com}, we receive instant ACKs but no SHs follow.
From our own domains targeted at the same rate as the Tranco domains, we almost always (99.9\%) receive an instant ACK.
When contacting six of our domains faster, we receive coalesced ACKs and ServerHellos more likely (7.5\%).

\begin{figure}
\includegraphics[width=\columnwidth]{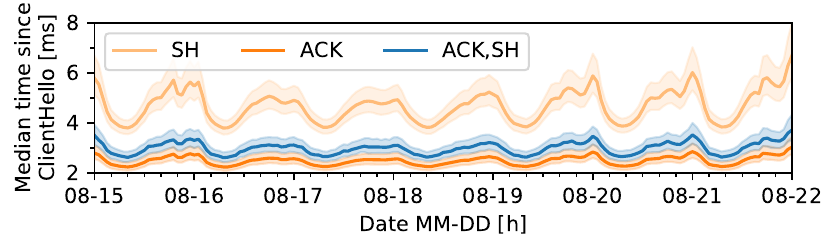}
\caption{Reception latency and 50~\% percentile interval of ACK and SH, either separately in sequential packets or coalesced ACK--SH from Cloudflare in Sao Paulo, BR. 
SH in coalesced messages arrive faster than separate SH.}
\label{fig:lin_cf-ack-sh-over-time-precalc}
\end{figure}

\autoref{fig:lin_cf-ack-sh-over-time-precalc} illustrates the behavior for Sao Paulo.
Coalesced ACKs and ServerHellos are received with similar delay compared to instant ACK, a strong indicator for caching.  
On median the instant ACK arrives after 2.1~ms (Sao Paulo, Hamburg), 2.4~ms (Los Angeles), and 2.6~ms before the SH (Hong Kong).
All vantage points show larger delays between IACK and ServerHello during local day time compared to the night (see \autoref{app:macroscopic-view}).
The observed delays are even smaller than the delays when connecting to all QUIC domains from the Tranco List (see \autoref{fig:CDF-ACK-SH-Delay}).
If the server had waited for the certificate from the certificate store instead, the PTO would have been inflated by 6.3 to 7.2~ms (up to 79\% of the median~RTT).
With this improvement and the measured RTTs, Cloudflare does not experience spurious retransmits and benefits from IACK.

\section{Discussion}
\label{sec:discussion}

\paragraphNoDot{Is instant ACK deployment only beneficial to CDNs with separate certificate store?}
No. There are multiple reasons for delays in server responses. 
Traversing a storage hierarchy, slow database lookups for configuration, high CPU utilization, and busy cryptography units are some examples that can cause belated responses from servers inflating the RTT estimate. 
In our setup, we found that responses with a (simple) certificate store operated on the frontend server may take longer than an IACK, mainly caused by the TLS signing function.

\paragraphNoDot{When does an instant ACK harm performance?}
If the RTT between frontend and backend server is larger than $\approx$3$\times$~RTT between client and frontend server, an instant ACK will lead to spurious retransmits.
Our analysis showed that these spurious retransmits may help when the server reaches the QUIC amplification limit, since the client packet (as a reply to the server instant ACK) provides the server additional sending budget.
Using a padded instant ACK to probe the path MTU, as Cloudflare implements, needs careful consideration, though, since this consumes additional amplification budget, which 
 can lead to an overall longer time until the handshake completes. 
If a server is not stalled by the anti-amplification limit, the probe packets add futile load to the network and server.

\paragraphNoDot{Can server operators rely on client implementations to correctly handle instant ACKs?}
No. 
We found that multiple implementations processed instant ACKs incorrectly, which added delays.
We work with the maintainers on a resolution of the issues.

\paragraphNoDot{How to improve instant ACK?}
Server and client behavior can be tuned to achieve better performance using instant ACK. 
When an instant ACK was received successfully but the ServerHello and additional packets of the handshake are lost, the server has to wait until its default PTO expires. 
Lowering this value is a trade-off between faster recovery from packet loss and inducing spurious retransmissions.
Alternatively, clients can cause the server to recover from this packet loss faster by retransmitting the ClientHello (when the PTO expires) instead of sending PING frames \cite{RFC-9002}.

\paragraph{Generalization to 0-RTT and Retry handshakes}
An instant ACK can also be used in case of 0-RTT handshakes to prevent PTO inflation.
Independently of the type of handshake, the client sends a ClientHello, which may trigger an instant~ACK.
The reception of a Retry packet causes the client to resend the ClientHello together with a token that proves its reception. 
Retry is implemented to counter resource exhaustion attacks. %
When a client receives the Retry packet it may use this packet as the first RTT estimate.
A subsequent instant ACK is still beneficial as it reduces RTT variation.

\paragraphNoDot{Do alternatives to instant ACK exist?}
Since the PTO initialization disregards the acknowledgment delay, IACK is the only mean to provide exact RTT estimates to the client at the beginning of a connection. 
A client could respect the acknowledgment delay at PTO initialization or reinitialize the PTO with the value of a subsequent, potentially undelayed RTT sample but the benefits are limited to subsequent exchanges or depend on correct server information (see \autoref{app:server-implementation} ).

\section{Conclusions}
\label{sec:conclusion}
In this paper, we discussed and empirically analyzed instant ACK, a standard compliant approach to improve the QUIC connection setup in common CDN deployments.
Instant ACK aims to better estimate the RTT between client and frontend server to adjust expiration timers more precisely---a fundamental challenge to reduce delays and prevent spurious retransmits seen in many transport protocols. Our measurements of popular domains revealed that Cloudflare, Akamai, Amazon, and Google deploy instant ACK in a significant share of domains. Depending on the scenario,  instant ACK proved beneficial or harmful.
At least one client implementation was negatively affected because it could not handle  instant ACKs.
We conclude that the protocol mechanics of QUIC challenge consistent performance improvement and  provide guidelines under which conditions instant ACK should be enabled in \autoref{app:iack-guideline}.

\label{lastpagebody}
\bibliographystyle{ACM-Reference-Format}
\bibliography{bibliography,rfcs,ids}

\appendix

\section{Ethics}
This work does not raise any ethical issues.

\section{Artifacts}
\label{app:artifcats}

We publish all artifacts of this paper, raw data, processing, and setup scripts. 

\paragraphNoDot{Data set:}
Qlog logs, packet captures, QScanner \cite{zbsja-io9ae-21} results

\paragraphNoDot{Run-time environment:} 
Python, Jupyter Notebooks

\paragraphNoDot{How much disk space is required?} 
\textasciitilde 600 GB

\paragraphNoDot{How much memory is required?}
256~GB

\paragraphNoDot{Publicly available?}
Yes

\paragraphNoDot{Archived?} 
Yes: \url{https://doi.org/10.5281/zenodo.13743301}

Our emulation uses Docker images pulled on 06.07.2024, except for \textit{mvfst}, which was pulled on 19.08.2024 to include a more recent Docker image compared to the start of our measurement.

\begin{table*}
\setlength{\tabcolsep}{4pt}
\caption{Deployment suggestions with and without packet loss. In the majority of scenarios, using instant ACK is advised.}
\label{tab:instant-ack-guidelines}
\begin{tabular}{p{6.5cm}cccc@{}}
\toprule
& \multicolumn{2}{c}{Loss} & \multicolumn{2}{c}{No Loss} \\\cmidrule(lr){2-3}\cmidrule(lr){4-5}
\multirow{2}{*}{\begin{tabular}[c]{@{}c@{}}Certificate size vs. amplification limit\end{tabular}} & \multirow{2}{*}{\begin{tabular}[c]{@{}c@{}}First server flight\\except first datagram\end{tabular}} & \multirow{2}{*}{\begin{tabular}[c]{@{}c@{}}Second client\\ flight\end{tabular}} & \multicolumn{2}{c}{Frontend-certificate store delay $\Delta t$} \\\cmidrule(lr){4-5}
&  &  & $\Delta t < $ 3 RTT (PTO) & $\Delta t \ge $ 3 RTT (PTO) \\
\midrule
(1) certificate size $\le$  anti-amplification limit & WFC & IACK & IACK & WFC \\
(2) certificate size $>$ anti-amplification limit & IACK & IACK & IACK & IACK \\
\bottomrule
\end{tabular}
\end{table*}

\section{Guidelines for Instant ACK Deployment}
\label{app:iack-guideline}
Certificate size and the delay between frontend and certificate store vary between different deployments. 
\autoref{tab:instant-ack-guidelines} summarizes when WFC and IACK should be preferred.
If the certificate size causes the handshake to exceed the anti-amplification limit, enabling instant ACK allows the server a higher sending budget and will improve performance.
Further, if the delay between frontend and certificate store is smaller than the client PTO (3$\times$ RTT), the client will react earlier to packet loss events. 
If instead, the server default PTO is important, \eg the first server flight, except the first datagram, is lost, an instant ACK does not provide the server with an RTT sample, which will slow server resends.
Last, if the frontend to certificate store delay is larger than the client PTO an instant ACK causes futile load on the server, which can decrease performance but is often tolerable by CDN deployments.
Optimally, servers should adjust the utilization of instant ACK depending on the expected certificate size and current frontend to certificate store~delay.

\section{ACK Delay: An Alternative to Instant ACK?}
\label{app:server-implementation}

In this section, we explain in detail why instant ACK cannot be completely replaced by using \texttt{ACK Delay}, a dedicated field in ACK~frames to encode the acknowledgment delay~\cite{RFC-9000}.
ACK Delay is not sufficient for three reasons.
First, the PTO initialization ignores the acknowledgment delay of the first RTT sample.
As such it would only help to reinitialize the PTO with the second sample.
Second, current server implementations often communicate an ACK Delay of 0.
Third, current CDN deployment challenge the use.

\paragraph{Limited support in server implementations}
We verify the support of ACK Delay by collecting the packet captures of successful handshake tests provided by the public QIR covering the 14 supported server implementations (\ie \textit{aioquic, go-x-net, haproxy, kwik, lsquic, msquic, mvfst, neqo, nginx, ngtcp2, picoquic, quic-go, quiche, quinn, s2n, xquic}), using a quic-go client from three subsequent days in August~2024.

\autoref{tab:server-ack-delays} indicates that 6 implementations report an ACK~Delay of 0~ms in the first packet.
\textit{msquic} does not send Initial and Handshake ACKs.
The remaining implementations report delays between 0.4~ms~(\textit{quinn}) and 15.2~ms~(\textit{s2n-quic}). 
The reported delay of \textit{s2n-quic} exceeds the RTT of the connection. 
11 implementations do not send an acknowledgment in the Handshake name space. 
3~implementations report an acknowledgment delay larger than~0~ms.

\begin{table}
\caption{Delay of the first acknowledgment received from server in the Initial and Handshake packet number space.}
\label{tab:server-ack-delays}
\begin{tabular}{lrrrrrr}
\toprule
 & \multicolumn{6}{c}{First ACK Delay [ms]} \\
 \cmidrule(lr){2-7}
 & \multicolumn{6}{c}{Repetition [index]} \\
 & 1 & 2 & 3 & 1 & 2 & 3 \\
 \cmidrule(lr){2-4} \cmidrule(lr){5-7}
Server & \multicolumn{3}{c}{Initial Packet} & \multicolumn{3}{c}{Handshake Packet} \\
\midrule
aioquic & 3.3 & 3.4 & 3.3 & - & - & - \\
go-x-net & 0.0 & 0.0 & - & - & - & - \\
haproxy & 1.0 & 1.0 & - & 0.0 & 0.0 & - \\
kwik & 0.0 & 0.0 & 0.0 & - & - & - \\
lsquic & 1.2 & 1.1 & 1.2 & 0.2 & 0.2 & 0.2 \\
msquic & - & - & - & - & - & - \\
mvfst & 0.8 & - & 0.7 & 0.2 & - & 0.1 \\
neqo & 0.0 & 0.0 & - & 0.0 & 0.0 & - \\
nginx & 0.0 & 0.0 & 0.0 & - & - & - \\
ngtcp2 & 0.0 & 0.0 & 0.0 & - & - & - \\
picoquic & 0.8 & 0.7 & 0.8 & - & - & - \\
quic-go & 0.0 & 0.0 & 0.0 & - & - & - \\
quiche & 1.4 & 1.4 & 1.5 & - & - & - \\
quinn & 0.4 & - & 0.4 & - & - & - \\
s2n-quic & 14.0 & 15.2 & 14.1 & - & - & - \\
xquic & 1.3 & 1.1 & 1.2 & - & 0.5 & 0.5 \\
\bottomrule
\end{tabular}
\end{table}

\begin{figure*}
\begin{subfigure}[t]{0.49\textwidth}
    \centering
    \includegraphics[width=\columnwidth]{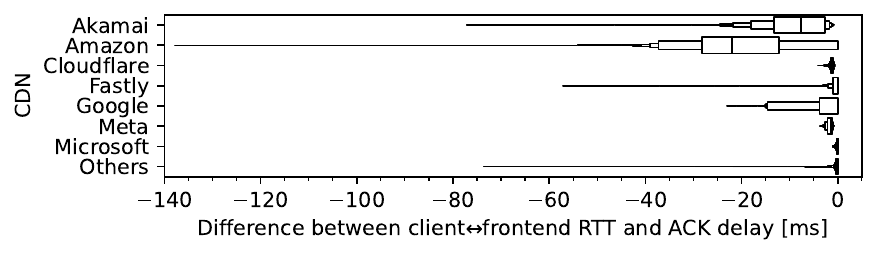}
    \caption{Coelesced ACK--SH}
    \label{fig:ack-delay-toplist-ack-sh}
\end{subfigure}
\begin{subfigure}[t]{0.49\textwidth}
    \centering
    \includegraphics[width=\columnwidth]{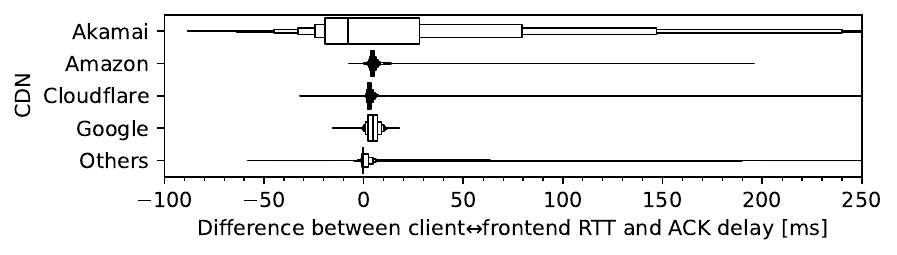}
    \caption{IACK}
    \label{fig:ack-delay-toplist-iack}
\end{subfigure}
\caption{Difference between RTT and acknowledgment delay. Coalesced ACK--SHs tend to carry an acknowledgment close to or exceeding the RTT. IACKs more frequently contain values lower than the RTT, allowing the client to correctly adjust the RTT sample.}
\label{fig:ack-delay-vs-rtt-toplist}
\end{figure*}

\paragraph{Challenges in the wild}
\autoref{fig:ack-delay-vs-rtt-toplist} shows that the acknowledgment delay exceeds the RTT in most responses with coalesced ACK--SH from Akamai (99.8\%), Amazon (87.3\%), Cloudflare (99.9\%), Fastly (60.5\%) and Meta (100\%) and Others (77.9\%).
For Google, we observe that this is the case in 34.8\%.
However, the difference between RTT and acknowledgment delay is less than 1ms for 99.8\% of the domains across all CDNs.
On the contrary, the majority of IACKs carry an acknowledgment delay larger than the RTT, except for Akamai and Others, where 61\% and 79.1\% are below that threshold.

According to the standard, in Initial packets clients may ignore the acknowledgment delay and in any packet number space, the acknowledgment delay should not be subtracted from the RTT sample, if the resulting value is lower than $min\_rtt$~\cite{RFC-9002}.
The consequence of sending such high acknowledgment delays in coalesced ACK--SH can lead to clients ignoring them or significantly underestimating the actual path RTT.
Current implementations challenge the use of this alternative.

\section{Client Implementation Details}
\label{app:client-implementation}

\paragraph{Default PTO}
RFC 9000 recommends a default PTO of 1~s. 
The majority of the analyzed implementations deviate significantly from the default to optimize recovery in case of packet loss.
\autoref{tab:initial-pto-second-flight} shows the designated default PTO by implementation.

\begin{table}
\caption{Initial PTO and UDP datagrams comprising the second client flight. Implementations chose lower initial PTOs than the recommended value of 1~s~\cite{RFC-9002} to improve recovery from packet loss. Due to packet coalescence the second client flight is sent in different UDP datagrams.}
\label{tab:initial-pto-second-flight}
\begin{tabular}{lcc}
\toprule
 & Default PTO & Second flight   \\
Implementation & [ms] & datagram(s) [index]  \\
\midrule
aioquic & 200 & 2,3,4 \\
go-x-net & 999 & 2,3,4 \\
mvfst & 100 & 2,3,4 \\
neqo & 300 & 2,3 \\
ngtcp2 & 300 & 2,3,4 \\
picoquic & 250 & 2,3,4,5 \\
quic-go & 200 & 2,3,4 \\
quiche & 999 & 2 \\
\bottomrule
\end{tabular}
\end{table}

\paragraph{Packet coalescence}
To analyze the performance impact of the instant ACK, we analyze QUIC connections with and without packet loss. 
QUIC allows coalescence of multiple QUIC packets in one datagram, \ie the information exchanged in one datagram varies. 
An unbiased experiment should compare equal information loss.
We analyze the content of each packet exchanged during the handshake and find that implementations coalesce QUIC packets differently but generally consistent, \eg the second client flight (see \autoref{fig:quic-connection-setup}) is transmitted in 1 to 4 datagrams.
We account for these differences by comparing loss of equal information. 
Hence, when analyzing loss of the second client flight, we respect the individual packet coalescence and compare loss of different datagrams. 
\autoref{tab:initial-pto-second-flight} specifies which datagrams this affects.
In case clients send additional packets (\eg when the default PTO expires), this static mapping no longer drops the entire second client flight (\cf \autoref{fig:ttfb-lscf-all-rtts}, 300~ms RTT).

\paragraph{Standard conformance and limitations of Qlog}
The transferability of our results to actual deployments, depends on the conformance of implementations to the standard and precision of information exposed in Qlog.
With respect to standard conformance, we find that \textit{aioquic} uses a different formula to calculate RTT variance.
Our calculation of PTOs is limited by the information exposed in Qlog.
Within these logs, we find that timestamps are provided with different resolutions, \ie µs, ms, and s, and \textit{neqo, mvfst} and \textit{picoquic} do not log RTT variance.
When RTT variance is not available, we calculate it from the sent and received packets~instead.

Next, we compare the number of RTT samples considered by implementations with the number of theoretically possible RTT samples, \ie reception of acknowledgments for packets that are newly acknowledged. 
We approximate the number of RTT samples of implementations by the number of changes in smoothed RTT and RTT variation. 
When implementations log the same information multiple times, we remove consecutive duplicates. 
This post-processing step might remove distinct samples represented by the same numbers.

\begin{figure}
    \includegraphics[width=\columnwidth]{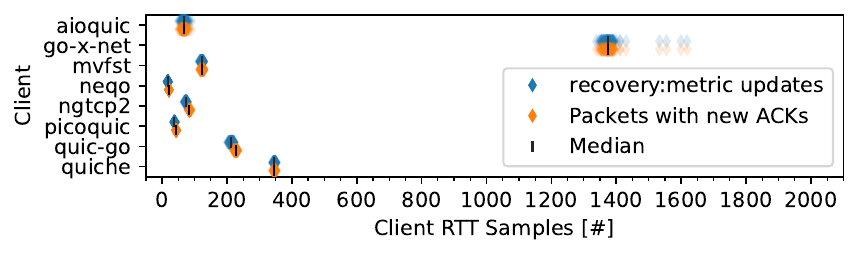}
\caption{Number of exposed RTT samples and newly acknowledging ACKs for 10~MB file transfer at 100~ms RTT,  WFC. 
    Due to different use of ACK-eliciting packets, \eg PING frames, implementations vary in the amount of RTT samples they can obtain. 
    They also expose different shares of the \textit{recovery:metric} updates.}
\label{fig:rtt_samples}
\end{figure}

\autoref{fig:rtt_samples} exhibits large differences in the number of RTT samples available to implementations.
Implementations may send PING frames, which causes more RTT samples.
Since the impact of the instant ACK on PTO declines and depends on the number of RTT samples, implementations with fewer RTT samples are influenced for a longer share of the connection.
Further, we observe that \textit{aioquic, go-x-net, mvfst}, and \textit{quiche} expose the maximum of PTO updates available, while \textit{neqo, ngtcp2, picoquic}, and \textit{quic-go} rely on a smaller fraction of the samples.
This indicates that an implementation \one does not utilize all received ACKs or \two does not log all metric~updates. 

\section{Details on Packet Loss Scenarios}
\label{app:interop-multiple-rtts}
We repeat all packet loss emulations with 1, 9, 20, 100, and 300~ms~RTT to ensure that our observations are valid across a wide range of network scenarios and compare the reception of the first payload byte (QUIC stream frame) after the loss event.

\begin{figure*}
\begin{subfigure}[t]{0.49\linewidth}
    \centering
    \includegraphics[width=\linewidth]{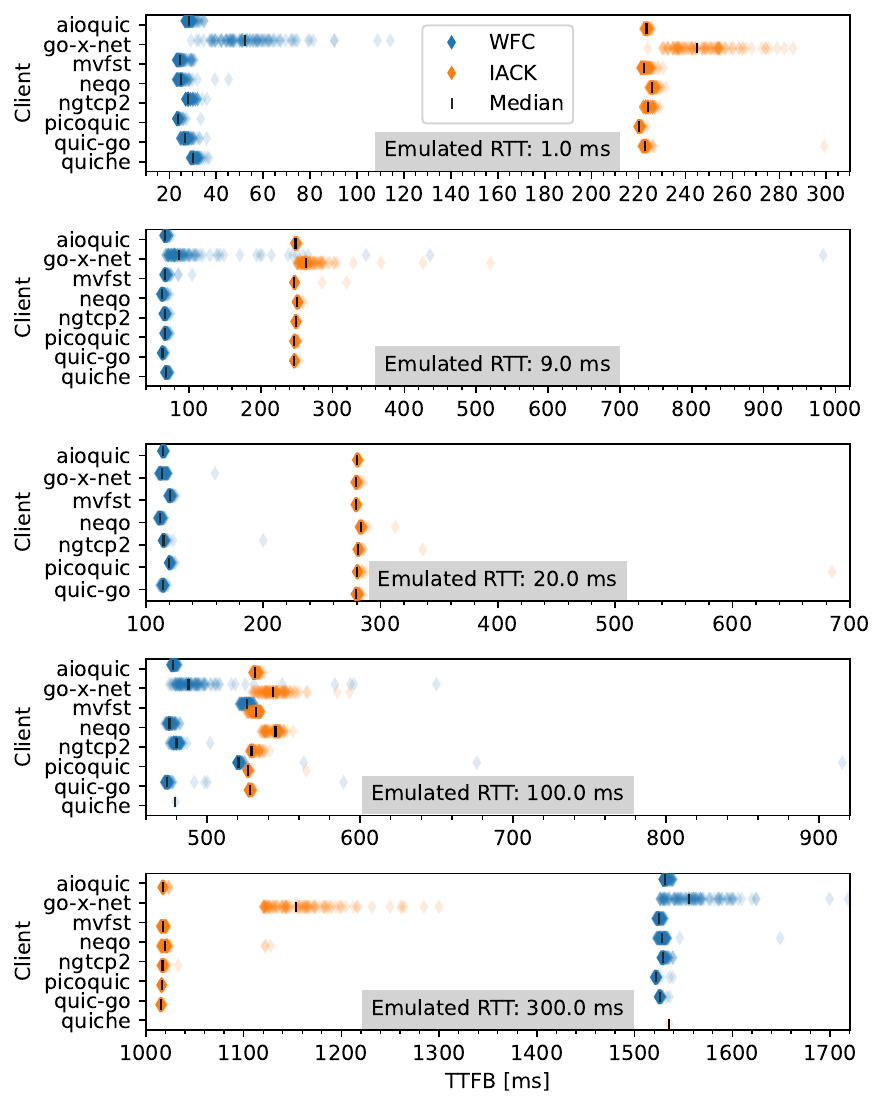}
    \caption{HTTP/1}
    \label{fig:ttfb-rfsf-goodput}
\end{subfigure}
\begin{subfigure}[t]{0.49\linewidth}
    \centering
    \includegraphics[width=\linewidth]{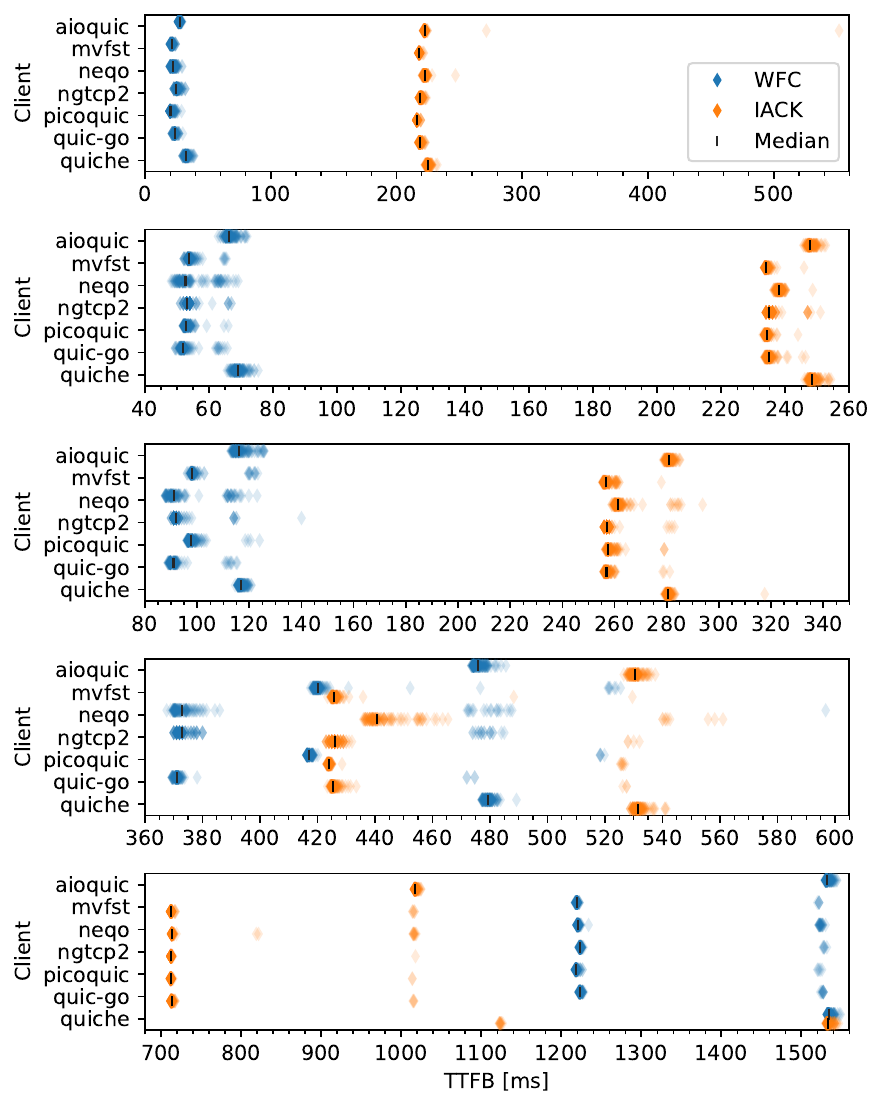}
    \caption{HTTP/3}
    \label{fig:ttfb-rfsf-http3}
\end{subfigure}
\caption{Time to First Byte of 10 KB file transfer at different RTTs under loss of packets 2 and 3 (IACK) and packet 2 (WFC) sent by the server. IACK prolongs the TTFB for all RTTs until the default PTO of the client is reached or until the PTO for the Handshake packet number space becomes relevant. Each row presents results from one emulated RTT.}
\label{fig:ttfb-rfsf-all-rtts}
\end{figure*}

\paragraph{Time to First Byte is increased when the remaining first server flight is lost}
\autoref{fig:ttfb-rfsf-all-rtts} shows that all QUIC connections until 100~ms RTT suffer from the server not holding an RTT estimate for the connection. 
At 100 ms RTT, the difference between IACK and WFC is reduced. 
In WFC, the TTFB is mainly influenced by the server receiving two RTT samples, \one in the Initial and \two in the Handshake packet number space, which reduces the variance and leads to setting an earlier PTO. 
For IACK, the server default PTO expires before the client sends probe packets (exceptions are \textit{mvfst, picoquic}). 
Only after an additional RTT, the client confirms the resent server Initials, which provides the server with a single RTT estimate that does not reduce the variance. 
A later Handshake PTO is set compared to WFC.
For \textit{mvfst} after 100 ms, in WFC the client default PTO expires and subsequently the server Handshake packet is acknowledged, which sets the Handshake PTO.
In IACK, \textit{mvfst} does not expire the default PTO after reception of an instant ACK.
Instead, the default PTO of the server expires and causes the client to acknowledge the retransmitted server packets, which provides the server with an RTT estimate.
Both lead to similar timings for resending the server Handshake packets.
For \textit{picoquic}, in WFC the acknowledgment of the server Handshake packet provides the server with an RTT estimate. 
In IACK, \textit{picoquic} does not send probe packets in response to the server instant ACK. Instead, the server PTO expires and picoquic computes the Handshake keys from the now received ServerHello and sends a Handshake PING packet, which causes the Handshake PTO to expire immediately.
Both behaviors lead to similar TTFB.

\begin{figure*}
\begin{subfigure}[t]{0.49\linewidth}
    \centering
    \includegraphics[width=\linewidth]{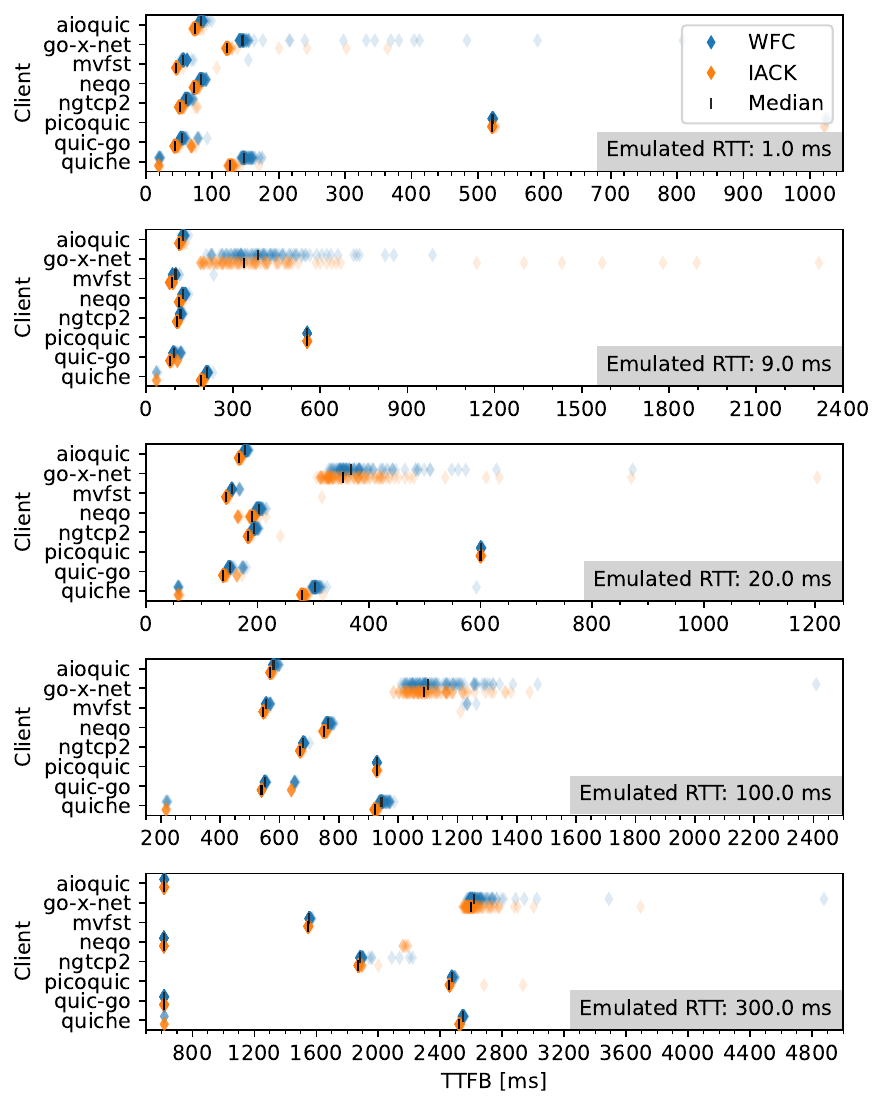}

    \caption{HTTP/1}
    \label{fig:ttfb-lscf-goodput}
\end{subfigure}
\begin{subfigure}[t]{0.49\linewidth}
    \centering
    \includegraphics[width=\linewidth]{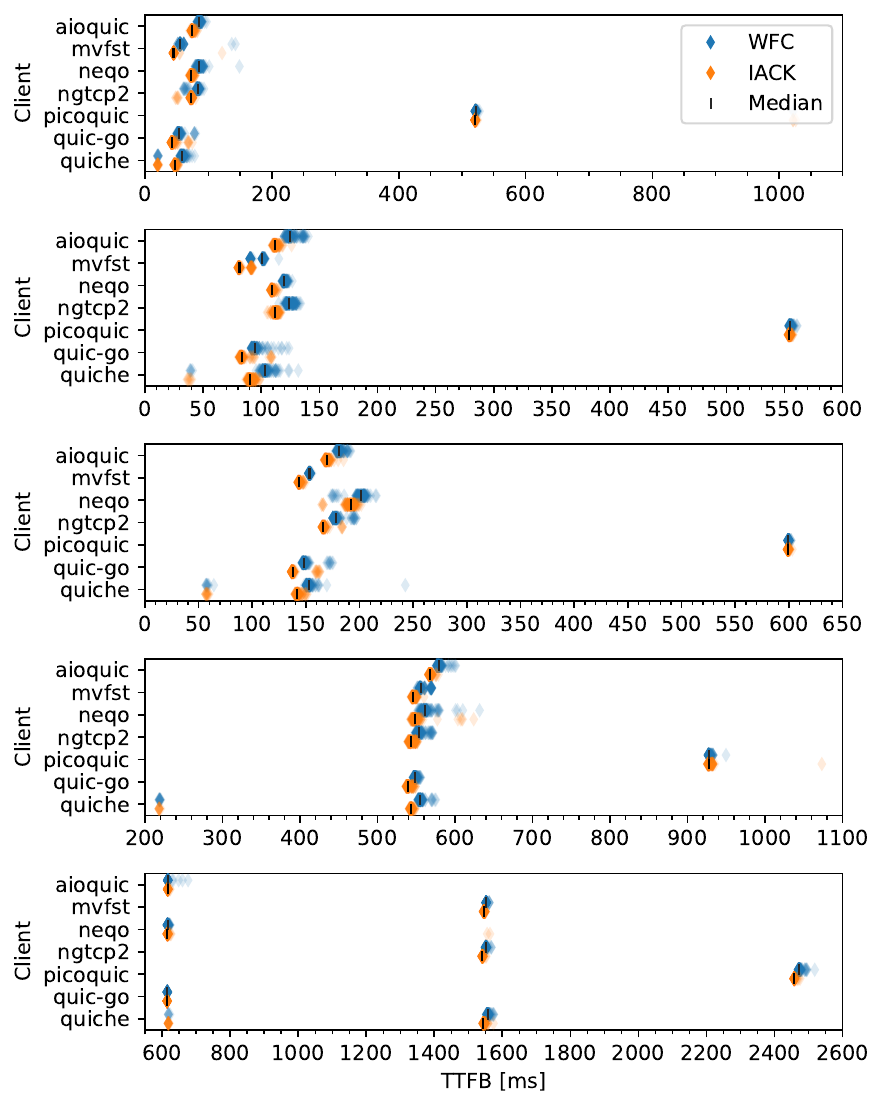}
    \caption{HTTP/3}
    \label{fig:ttfb-lscf-http3}
\end{subfigure}
\caption{Time to First Byte of 10 KB file transfer at different RTTs under loss of the entire client second flight. IACK improves the TTFB. Each row presents results from one emulated RTT.}
\label{fig:ttfb-lscf-all-rtts}
\end{figure*}

At 300~ms RTT, IACK outperforms WFC due to restarting the PTO timer with a larger than default PTO after reception of an acknowledgment. 
In the case of IACK, the server only receives acknowledgments in the Initial packet number space, while in the case of WFC, the first packet contains a Handshake packet, which is acknowledged by the client.
On reception of the Handshake ACK, the server sets a PTO based on the larger RTT for the outstanding Handshake packets, which in consequence delays the TTFB. 
The PTO is then larger than the default PTO of our server implementation, which leads to the differences.
For \textit{quiche}, at all RTTs the duplicate CID retirement issue caused abortion of the measurement immediately or after up to one successful exchange.
Please note that the improvement of $\approx$200 ms originates from the default server PTO, a higher default server PTO will lead to a different advantage of WFC over IACK.
Similar behavior is observed for HTTP/3.

\paragraph{Time to First Byte is reduced when the second client flight is lost}
\autoref{fig:ttfb-lscf-all-rtts} shows a general improvement of IACK over WFC for all RTTs.
\textit{picoquic} does not account for the shorter RTT estimate with IACK and relies solely on its default PTO. 
The PTO timer is reset after a retransmission from the server is received, due to the server default PTO expiring in IACK and WFC.
For \textit{quiche}, two kinds of behaviors are observable, the lighter diamonds at an earlier time originate from sending two packets instead of one packet for the client second flight.
Only one packet is dropped by our emulation (see \autoref{tab:initial-pto-second-flight}). 
The server then already holds the client request and is able to serve it sooner.
At 300~ms RTT for \textit{aioquic, mvfst, neqo, ngtcp2} and \textit{quic-go} the default PTO expires before any further packet from the server is processed and the second client flight can be sent. 
Instead, the ClientHello is deemed lost and resent. 
Our static mapping of the second client flight then drops the probe packets, which cause the HTTP request to arrive earlier (see \autoref{tab:initial-pto-second-flight}).
\textit{quiche} and \textit{go-x-net} use a large default PTO and are thus not impacted by probe packet drop instead of second client flight drop.
Similar behavior can be observed for our HTTP/3 measurement.

\section{Macroscopic View}
\label{app:macroscopic-view}

\begin{table}
\caption{AS numbers used for CDN inferences.}
\label{tab:cdn-as}
\begin{tabular}{ll}
\toprule
 CDN & AS numbers \\
\midrule
Akamai & 16625, 20940 \\
Amazon & 14618, 16509 \\
Cloudflare & 13335, 209242 \\
Fastly & 54113 \\
Google & 15169, 396982 \\
Meta & 32934 \\
Microsoft & 8075 \\
\bottomrule
\end{tabular}
\end{table}

\paragraphNoDot{Which QUIC implementation is deployed by a CDN?}
Different behavior of CDNs may originate from different implementations (\eg Cloudflare deploys \textit{quiche} with IACK, while Meta deploys \textit{mvfst} without IACK). 
Zirngibl \etal~\cite{zgssc-qhfqd-24} show that the order of QUIC transport parameters and the content of \texttt{CONNECTION\_CLOSE} frames can be used to identify server implementations.
Identifying different QUIC stacks is beyond the scope of this work, but applying such a method could be part of future work.

\paragraph{Assignment of IP addresses to on-net CDN deployments}
CDN hosted domains are inferred from their IP~addresses mapped to origin ASes gained from route announcements on August 7, 2024.
To account for CDNs operating multiple ASes, we assign multiple AS numbers to one CDN according to \autoref{tab:cdn-as}.

\begin{figure}
    \includegraphics[width=\columnwidth]{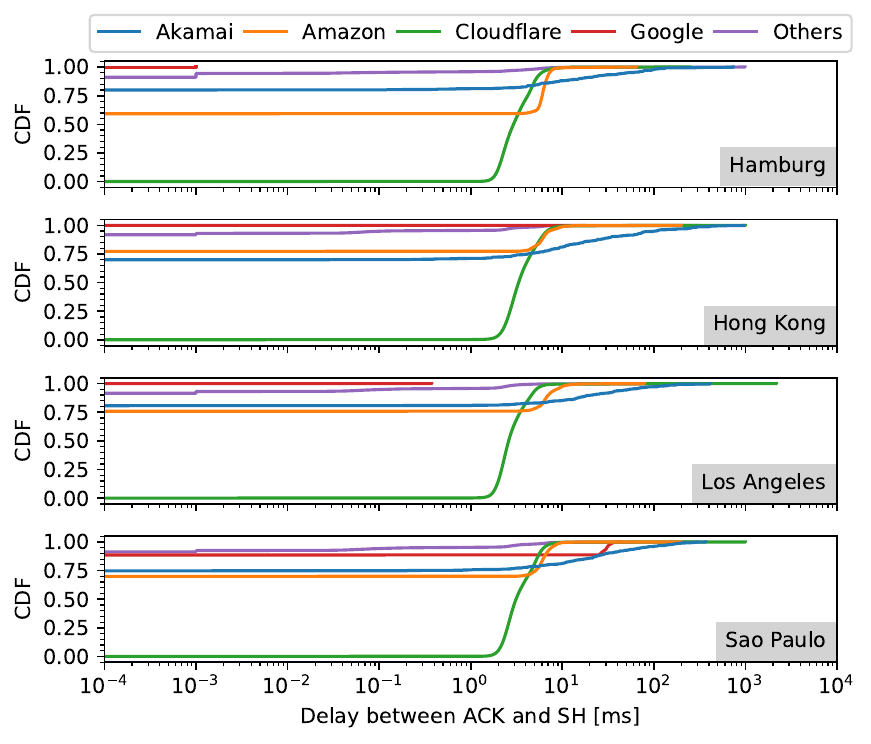}
    \caption{Delay between reception of the first ACK and subsequent ServerHello (SH) from our four vantage points for domains on the Tranco Top 1M. IACK performance is similar across locations.}
    \label{fig:toplist-locations}
\end{figure}

\paragraph{CDN deployment variations}
\autoref{fig:toplist-locations} exhibits homogeneous delays between instant ACK and later ServerHello from Akamai, Amazon, Cloudflare, and Others from all four measurement locations. 
Google IACK-enabled servers are only significantly reachable from Sao Paulo.
Otherwise, implementations vary in the share of IACK deployment by up to 8.3\% between different days from the same location.

\begin{figure}
  \includegraphics[width=\columnwidth]{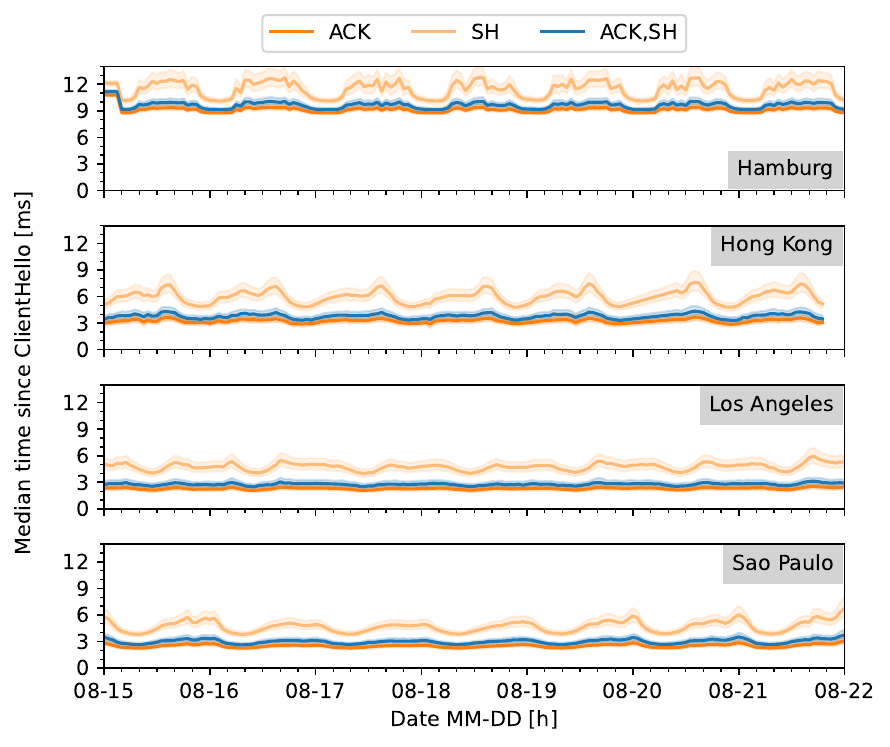}
  \caption{Time between request and response from Cloudflare servers from the measurement locations with 50 \% percentile interval. The gaps in the measurements from Hong Kong are caused by a misconfiguration of our nodes.}
  \label{fig:cloudflare-ack-sh-intervals-global}
\end{figure}

\paragraph{Cross-verification of Cloudflare IACK deployment}
For cross-verification, we performed the measurement from a European university network~(Hamburg, DE) and Google Cloud in North America~(Los Angeles, US), South America~(Sao Paulo, BR) and Asia~(Hong Kong, HK) in August~2024.
\autoref{fig:cloudflare-ack-sh-intervals-global} shows the median time between the request and response packet from Cloudflare servers at the same location as our probe. 
At all locations the coalesced ACK--SH is faster than the separated ServerHello.
Gaps in the measurement from our vantage point are caused by host maintenance temporarily deactivating our virtual machine.

\begin{figure}[h!]
\includegraphics[width=1.00\linewidth]{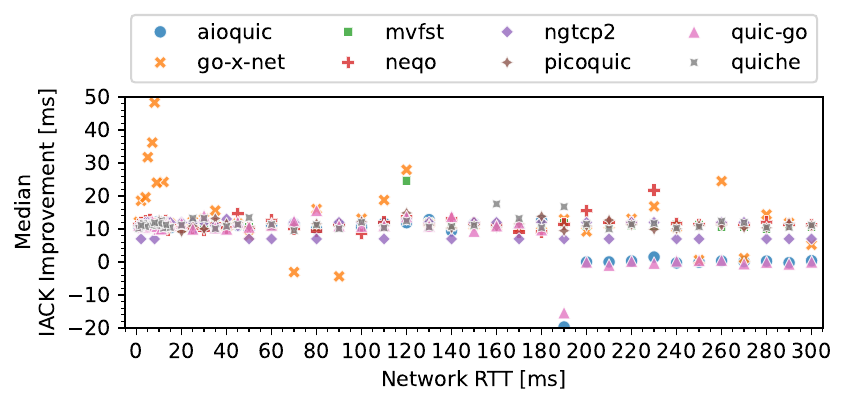}
\caption{Improvement of the first PTO, based on recovery metric updates in Qlog.
The variance is calculated from the logged packet receptions, if it is not provided by the implementation.
}
\label{fig:iack-improvement-factors-metrics}
\end{figure}

\paragraph{IACK provides a consistent PTO improvement across RTTs}
\autoref{fig:iack-improvement-factors-metrics} shows the median PTO improvement of IACK over WFC derived from the first smoothed RTT and RTT variance information provided by the implementations.
If the variance is not provided, we calculate it based on packet reception times.
Implementations exhibit similar PTO improvements across all RTTs.
\textit{go-x-net} partially initializes the PTO incorrectly, which also influence the median difference.
Beginning at 190~ms RTT, \textit{quic-go} and \textit{aioquic} only log RTT estimates after multiple RTTs.
This correlates with their default PTO expiring at that same time.

\end{document}